\documentclass[prb,twocolumn,aps,superscriptaddress,floatfix]{revtex4-2}
\usepackage{amsmath}
\usepackage{amssymb}
\usepackage{bm}
\usepackage{graphicx}
\usepackage{color}
\usepackage{hyperref}
\usepackage{verbatim}% Long comments

\begin{document}

\title{Graphene on a ferromagnetic substrate: instability of the electronic
liquid}

\author{D.N. Dresviankin}
\affiliation{Institute for Theoretical and Applied Electrodynamics, Russian
Academy of Sciences, 125412 Moscow, Russia}
%\affiliation{Moscow Institute of Physics and Technology, Dolgoprudny, 141700, Russia}

\author{A.V. Rozhkov}
\affiliation{Institute for Theoretical and Applied Electrodynamics, Russian
Academy of Sciences, 125412 Moscow, Russia}

\author{A.O. Sboychakov}
\affiliation{Institute for Theoretical and Applied Electrodynamics, Russian
Academy of Sciences, 125412 Moscow, Russia}

\begin{abstract}
We previously show
[JETP Letters, {\bf 114}, 763 (2021)]
that a graphene sample placed 
on a ferromagnetic substrate demonstrates a cooperative magnetoelectronic
instability. The instability induces a gap in the electronic spectrum and a
canting deformation of the magnetization near the graphene-substrate
interface. In this paper we prove that the interaction between the
electrons in graphene strongly enhances the instability. Our estimates
suggest that in the presence of even a moderate interaction the instability
can be sufficiently pronounced to be detected experimentally in a realistic
setting.
\end{abstract}

\date{\today}

\maketitle

\section{Introduction}

%Graphene-based spintronics remains an active area of
%research~\cite{Han2014}. 
A graphene sheet placed on a ferromagnetic insulating 
substrate~\cite{graph_fm_hetero2015exper,Wei2016,graphene_YIG2015exper,
graphene_YIG2016exper}
may be viewed as a prototypical
graphene-spintronic~\cite{Han2014}
device. In such a heterostructure, due to the magnetic proximity effect,
the spins of the graphene electrons are polarized. This polarization is
accompanied by emergence of the Fermi surfaces in both graphene valleys,
turning the graphene, which is a semi-metal in its pristine form, into a
self-doped metal. Experimental study of such a graphene-based magnetic
metal have been reported in
Ref.~\onlinecite{Wei2016},
for example. As for theoretical studies dedicated to this, and similar,
setups, one can mention
Refs.~\cite{Zollner2016,spin_valve2018theory,bilayer_fm2018theory}.

In our recent
paper~\cite{our_JETP_Lett}
we have demonstrated that, at low temperature, a graphene sample placed in
an insulating ferromagnetic substrate experiences a cooperative
magnetoelectronic instability: due to the Fermi surface nesting, the
perfect homogeneous ferromagnetic polarizations of both the graphene and
the substrate experience canting, while the gap opens in the
single-electron spectrum of the graphene. As a result of the instability,
the magnetic metal becomes a magnetic semiconductor (that is, an insulator
with a small gap).

To offer an intuitive and clear picture of the mechanism behind the
instability, the presentation of
Ref.~\onlinecite{our_JETP_Lett}
was made intentionally simple. An unfortunate downside of this approach is
its diminishing reliability. The purpose of the present paper is to develop
a more accurate description of the instability under study.

The simplifications incorporated into the theoretical model of
Ref.~\onlinecite{our_JETP_Lett}
are of two sorts: (i)~all single-electron states whose energies 
$\varepsilon_{\bf k}$
lie too far from the Fermi energy were neglected (specifically, all states
with
$|\varepsilon_{\bf k}|$
exceeding
$\sim~1$\,eV
were omitted),
(ii)~the electron-electron interaction in graphene was ignored. Of these
two, assumption~(i) is purely technical: its only role was to justify the
use of linear density of states (DOS) for graphene. It can be amended
without introducing new concepts to the theoretical formalism of
Ref.~\onlinecite{our_JETP_Lett}.
The situation with~(ii) is more complex, and requires more advanced
theoretical apparatus. 

The present paper addresses both (i) and (ii). Namely, we use the
tight-binding DOS to account for all
$p_z$
electronic states. Most importantly, we explicitly include a (Hubbard-like)
interaction into the model. The interaction is then treated at the mean
field level. 

We will see that both modifications to the formalism act to increase the
instability strength and the gap value relative to the expressions derived
in
Ref.~\onlinecite{our_JETP_Lett}.
This makes it easier to argue that the instability may be observed in an
experiment under realistic conditions.

Beside this, our formalism allows us to reveal the
interaction-strength-driven crossover between magnetoelectronic instability
(which relies on cooperation between the magnetic and electronic
subsystems) to a more common spin-density wave (SDW) instability of purely
electronic origin.

The paper is organized as follows. In
Sec.~\ref{sec::geometry}
we describe the geometrical aspects of the studied heterostructure. The
model Hamiltonian is introduced in
Sec.~\ref{sec::model}.
The magnetoelectronic instability is discussed in
Sec.~\ref{sec::instability0},
while the interaction effects are discussed in
Sec.~\ref{sec::interaction}.
Section~\ref{sec::discussion}
is reserved for the discussion of the paper's main results and the
conclusions. Technically involved derivations are relegated to Appendices.

\begin{figure}[t]
\includegraphics[width=0.45\columnwidth]{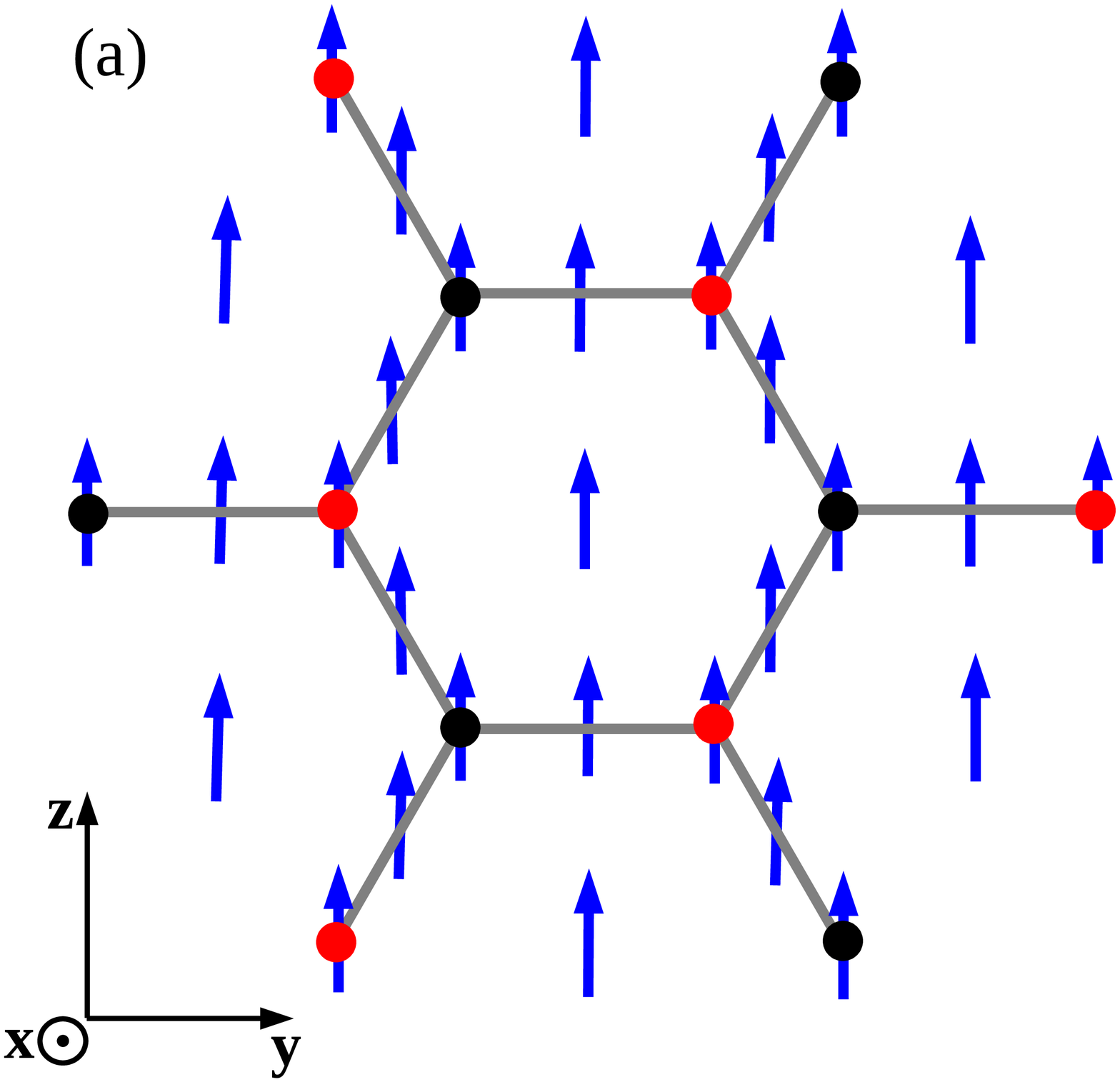}
\includegraphics[width=0.45\columnwidth]{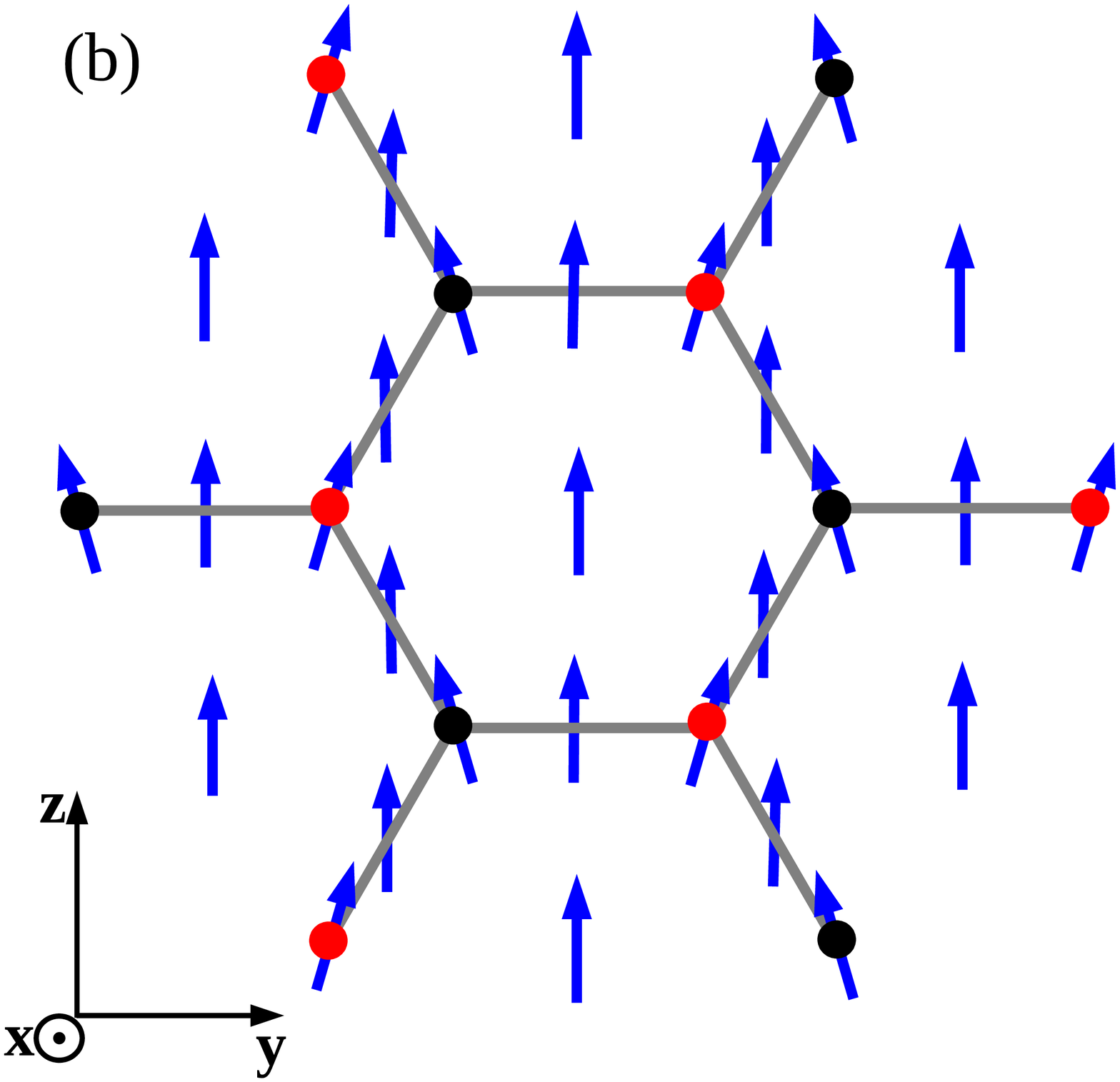}
\caption{Schematic representation of a graphene sample on a ferromagnetic
substrate. The orientation of the axes is shown in the lower left corner.
The origin is at the center of the regular hexagon. Carbon atoms of
graphene are represented by black (sublattice B) and red (sublattice A)
circles. The solid (grey) lines connecting the atoms show carbon-carbon
chemical bonds. There is a ferromagnetic substrate under graphene. Blue
arrows represent local magnetization on the surface of the ferromagnet.
Panel~(a) corresponds to the case of perfect magnetization, see
Eq.~(\ref{eq::M_FM}).
Panel~(b) represents the canted state. In this case, magnetization
projection
$M_y$
varies in space periodically, see
Eqs.~(\ref{spatialmagn})
and~(\ref{eq::canted_condition}).
We consider only those canting deformations, for which 
$M_y$
has opposite signs beneath the atoms belonging to different sublattices, in
agreement with
condition~(\ref{eq::canted_condition}).
One can easily recognize in panel~(b) that 
$M_y > 0$
beneath the atoms of $A$ sublattice, and
$M_y < 0$
for $B$ sublattice.
%%%%%%%%%%%%%%%%%%%%%%%%%%%%%%%%%%%%%%%%%%%%%%%%%% 
\label{fig::spec_numer_app}
%%%%%%%%%%%%%%%%%%%%%%%%%%%%%%%%%%%%%%%%%%%%%%%%%% 
}
\end{figure}

\section{Geometry considerations}
%%%%%%%%%%%%%%%%%%%%%%%%%%%%%%%%%%%%%%%%%%%%%%%%%%
\label{sec::geometry}
%%%%%%%%%%%%%%%%%%%%%%%%%%%%%%%%%%%%%%%%%%%%%%%%%%

The main object of our study is a graphene sample placed on a ferromagnetic
substrate. Below we assume that the graphene lies in the
$Oyz$~plane, while the
$Ox$-axis
is perpendicular to the substrate and directed away from it surfaces,
see
Fig.~\ref{fig::spec_numer_app}\,(a).
This is a ``non-canonical" orientation of the coordinate system (usually it
is assumed that graphene is located in the
$Oxy$~plane). However, our
choice makes description of the magnetic subsystem more conventional, since
it will allow us to use
$Oz$
as the spin quantization axis. Symbols
$\hat{e}_{x,y,z}$
denote the unit vectors in the direction of the corresponding axes. To
specify a point on the substrate surface we will use two-dimensional
vectors
${\bf R} = y\hat{e}_y + z\hat{e}_z$
while a point inside the substrate is specified by vector
${\bf r} = x\hat{e}_x + y\hat{e}_y + z\hat{e}_z$,
$x<0$,
with
$x=0$
being the substrate surface.

The graphene lattice is described by elementary translation vectors 
\begin{eqnarray}
{\bf a}_{1,2} = \frac{\sqrt{3}}{2} a_0
\left( \sqrt{3} \hat{e}_y \pm  \hat{e}_z \right),
\end{eqnarray} 
where 
$a_0$
is the distance between neighboring carbon atoms in graphene. Lattice sites
coordinates on the sublattices
$A$
and
$B$
are given by the vectors 
\begin{eqnarray}
{\bf R}_{A, B} = {\bf a}_1 n_1 + {\bf a}_2 n_2 \mp a_0 \hat{e}_y, \quad
n_{1,2} \in \mathbb{Z}.
\end{eqnarray} 
In
Fig.~\ref{fig::spec_numer_app}
lattice sites corresponding to different sublattices are depicted in
different colors.

Finally, let us remind that the Brillouin zone in graphene has the shape of
a regular hexagon, and the expression
\begin{eqnarray}
{\bf b}_{1,2} = \frac{2 \pi}{3 a _0} (\hat{e}_y \pm \sqrt{3} \hat{e}_z)
\end{eqnarray} 
specifies the reciprocal lattice vectors for graphene.

\section{Model for the graphene on a ferromagnetic substrate}
%%%%%%%%%%%%%%%%%%%%%%%%%%%%%%%%%%%%%%%%%%%%%%%%%%
\label{sec::model}
%%%%%%%%%%%%%%%%%%%%%%%%%%%%%%%%%%%%%%%%%%%%%%%%%%

Our model Hamiltonian describing electrons on the hexagonal lattice of
graphene placed in contact with the ferromagnetic substrate reads
\begin{eqnarray}
%%%%%%%%%%%%%%%%%%%%%%%%%%%%%%%%%%%%%%%%%%%%%%%%%%
\label{eq::H_model}
%%%%%%%%%%%%%%%%%%%%%%%%%%%%%%%%%%%%%%%%%%%%%%%%%% 
\hat{H} = 
\hat{H}_0 + \hat{H}_{\rm Z} + \hat{H}_{\rm HB}.
\end{eqnarray}
Here
$\hat{H}_0$
is the usual tight-binding Hamiltonian of the graphene
\begin{eqnarray}
	\hat{H}_0 = -t \sum_{\langle \mathbf{R}_A, \mathbf{R}_B \rangle, \, \sigma} 
		\hat{d}_{ \mathbf{R}_{A} \sigma}^{\dag} 
		\hat{d}_{\mathbf{R}_{B} \sigma}^{\vphantom{\dagger}} 
		+ {\rm H.c.}
\end{eqnarray} 
In this expression the summation runs over the nearest-neighbor pairs
${\langle \mathbf{R}_A, \mathbf{R}_B \rangle}$
and spin projection $\sigma$.
Symbol $t$ denotes the hopping integral (for calculations one can use
$t \approx 2.7$\,eV).

Interaction between the electrons is described by the Hubbard term
\begin{eqnarray}
\hat{H}_{\rm HB}
=
U \sum_{{\bf R}_\alpha }
	\hat{n}_{{\bf R}_\alpha \uparrow} \hat{n}_{{\bf R}_\alpha \downarrow},
	\\
	\text{where}\quad
	\hat{n}_{{\bf R}_\alpha \sigma} = \hat{d}_{ \mathbf{R}_{\alpha} 
	\sigma}^{\dag} \hat{d}_{\mathbf{R}_{\alpha} \sigma}^{\vphantom{\dagger}},
\end{eqnarray} 
is the electron number operator and 
$\alpha = A, B$
is the sublattice index. Finally, the Zeeman term 
$\hat{H}_{\rm Z}$
induced due to the proximity to magnetic substrate equals to
\begin{eqnarray}
\hat{H}_{\rm Z}
=
\sum_{{\bf R}_\alpha}
	{\bf h}_{{\bf R}_\alpha} \cdot \hat{\bf S}_{{\bf R}_\alpha}.
\end{eqnarray}
In this formula the electron spin operator
$\hat{\bf S}_{{\bf R}_\alpha}$
is defined by a familiar expression
\begin{eqnarray}
\hat{S}_{{\bf R}_\alpha}^{j} = \left( \begin{matrix}
    \hat{d}_{ \mathbf{R}_{\alpha} \uparrow}^{\dag} 
	\hat{d}_{ \mathbf{R}_{\alpha} \downarrow}^{\dag} 
	\end{matrix} \right) 
	\hat{\sigma}^{j}
	\left( \begin{matrix}
    \hat{d}_{ \mathbf{R}_{\alpha} \uparrow}^{\vphantom{\dagger}} 
    \\
	\hat{d}_{ \mathbf{R}_{\alpha} \downarrow}^{\vphantom{\dagger}} 
	\end{matrix} \right),
\end{eqnarray} 
where
$\hat{\sigma}^{j}$
are the Pauli matrices. Quantity
${\bf h}_{{\bf R}_\alpha}$
in
$\hat{H}_{\rm Z}$
is the exchange field experienced by electrons on sublattice $\alpha$, on
unit cell
${\bf R}$.
We will assume that there is simple proportionality relation between
${\bf h}_{{\bf R}_\alpha}$
and local substrate magnetization
${\bf M} ({\bf r})$
\begin{eqnarray}
%%%%%%%%%%%%%%%%%%%%%%%%%%%%%%%%%%%%%%%%%%%%%%%%%%
\label{eq::proximity}
%%%%%%%%%%%%%%%%%%%%%%%%%%%%%%%%%%%%%%%%%%%%%%%%%% 
{\bf h}_{{\bf R}_\alpha} = \tau {\bf M} (0, {\bf R}_\alpha ).
\end{eqnarray}
In this relation, coefficient $\tau$ represents the strength of the
magnetic proximity effect. 
Equation~(\ref{eq::proximity})
implies that the Zeeman field at a specific carbon atom is proportional to
the substrate magnetization directly beneath this atom.

\begin{figure}
    \centering
    \includegraphics[width=0.95\columnwidth]{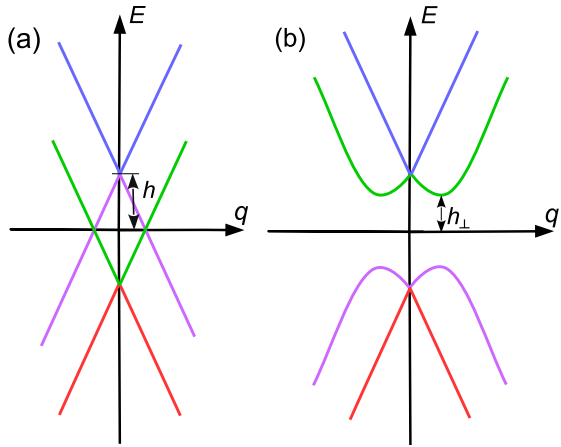}
    \caption{Schematic representation of the low-energy electron dispersion
for graphene on a ferromagnetic substrate. The horizontal axis represents
the two-dimensional momentum space. Energy is shown on the vertical axis.
The origin corresponds to the Dirac point (the spectrum is
identical near both Dirac points). For electronic states with spin 
$\sigma = \uparrow$
($\sigma = \downarrow$),
the apex of the Dirac cone is shifted by $h$ (shifted by $-h$). The
dispersion for the perfectly homogeneous magnetization is plotted in~(a).
The band shown by the blue (red) line is completely empty (filled). The
purple and green lines correspond to the bands crossing the Fermi level.
These bands form an electron and a hole Fermi surface. The electron and
hole Fermi surfaces coincide, exhibiting the perfect nesting with zero
nesting vector.
The insulating state with the canted magnetization is presented in~(b). The
filled bands (red and purple curves) contain mixed electronic states
resulting from the hybridization of electron and hole states. The empty
bands are shown in green and blue. The spectrum possesses a gap of
$2h_\perp$.
%%%%%%%%%%%%%%%%%%%%%%%%%%%%%%%%%%%%%%%%
\label{fig::dispersion}
%%%%%%%%%%%%%%%%%%%%%%%%%%%%%%%%%%%%%%%%
    }
\end{figure}

For vanishing interaction 
$U=0$
and homogeneous ferromagnetic magnetization
\begin{eqnarray}
%%%%%%%%%%%%%%%%%%%%%%%%%%%%%%%%%%%%%%
\label{eq::M_FM}
%%%%%%%%%%%%%%%%%%%%%%%%%%%%%%%%%%%%%%
{\bf M} = M (0, 0, 1), \quad M >0,
\end{eqnarray} 
the Hamiltonian
$\hat{H}$
reads
\begin{eqnarray} 
%%%%%%%%%%%%%%%%%%%%%%%%%%%%%%%%%%%%%%%%%%%%%%%%%%
\label{eq::H_noninteract}
%%%%%%%%%%%%%%%%%%%%%%%%%%%%%%%%%%%%%%%%%%%%%%%%%% 
\hat{H} = \sum _{\mathbf{q}}
    \hat\Phi_{\bf q}^\dag
    \hat{\cal H}_{\bf q}^{\vphantom{\dag}}
    \hat\Phi_{\bf q}^{\vphantom{\dag}},
\end{eqnarray}
where
$\hat\Phi_{\bf q} =
\left( 
        \hat{d} _{\mathbf{q} A \uparrow},
        \hat{d} _{\mathbf{q} B \uparrow},
        \hat{d} _{\mathbf{q} A \downarrow},
        \hat{d} _{\mathbf{q} B \downarrow}
\right)^T$
is a bi-spinor operator corresponding to states with quasimomentum
$\bf q$.
In this expression
$\hat{d}_{{\bf q} \alpha \sigma}$
is annihilation operator of electron with quasi-momentum
${\bf q}$,
located on sublattice
$\alpha$
with spin
$\sigma = \uparrow, \downarrow$.
Matrix
$\hat{\cal H}_{\bf q}$
reads
\begin{eqnarray}
%%%%%%%%%%%%%%%%%%%%%%%%%%%%%%%%%%%%%%%%%%%%%%%%%% 
\label{Hamiltonian0}
%%%%%%%%%%%%%%%%%%%%%%%%%%%%%%%%%%%%%%%%%%%%%%%%%% 
\hat{\cal H}_{\bf q}
=
\left( \begin{matrix}
    h & -t f_{\mathbf{q}} & 0 & 0 \\
    -t f _{\mathbf{q}} ^{*} & h & 0 & 0 \\
    0 & 0 & -h & -t f _{\mathbf{q}} \\
    0 & 0 & -t f _{\mathbf{q}} ^{*} & -h 
    \end{matrix} \right),
\end{eqnarray}
where local Zeeman field
$h$
equals to
$h = \tau M$,
and function
$f _{\mathbf{q}}$
can be expressed as
\begin{eqnarray}
f_{\bf q}
=
\left[
    1 + 2 \exp{\left( \frac{3 i a _0 q_y}{2} \right)
	    \cos{\left( \frac{\sqrt{3} a_0 q_z}{2} \right)}}
\right].
\end{eqnarray}
Diagonalizing 
$\hat{\cal H}_{\bf q}$
we derive electron dispersion for graphene on ferromagnetic substrate 
\begin{eqnarray}
%%%%%%%%%%%%%%%%%%%%%%%%%%%%%%%%%%%%%%%%%%%%%%%%%%
\label{eq::dispersion0}
%%%%%%%%%%%%%%%%%%%%%%%%%%%%%%%%%%%%%%%%%%%%%%%%%%
\varepsilon^{(1,2,3,4)}_{\bf q} = \pm h \pm t |f_{\bf q}|.
\end{eqnarray}
This expression demonstrate that in the presence of the magnetic substrate
the electronic structure is composed of four non-degenerate bands. Two of
them cross Fermi level, forming circular Fermi surface around each
non-equivalent Dirac points. Resultant dispersion is shown in
Fig.~\ref{fig::dispersion}.

Note that the Fermi surface is formed without external doping. In other
words, the Zeeman field leads to self-doping effect: the electrons leave
the
$\sigma = \uparrow$
band, accumulating instead in the
$\sigma = \downarrow$
band, generating two Fermi surface components. Due to this spin-dependent
mechanism, the emerged Fermi surfaces are not spin-degenerate, unlike the
situation in an ordinary metal. 

Another important property of this electronic structure is the nesting of
the Fermi surface. That is, the hole Fermi surface (formed by
single-particle states with spin
$\sigma = \uparrow$)
coincides with the electronic Fermi surface (formed by single-particle
states with spin
$\sigma = \downarrow$).
It is well known that a Fermi surface with nesting loses its stability when
collective effects are taken into account. Indeed, the nesting is one of
the major ingredients underpinning the mechanism of the magnetoelectronic
instability discussed in our
Ref.~\onlinecite{our_JETP_Lett}.

Finally, let us comment that inclusion of repulsive interaction
$U > 0$
into the model introduces several modifications to the non-interacting
electronic state described in the previous paragraphs. For one, finite
electron-electron interaction renormalizes Zeeman susceptibility. This
effect is weak, as we will see below. More importantly, however, the
interaction may induce a transition into an SDW phase (see, for example,
Ref.~\onlinecite{sdw_inplane_magnet2007theor}
for a related discussion). Our investigation will show that, at low
temperature, two incipient instabilities (SDW and magnetoelectronic)
interact and mutually enhance each other.

\section{Magnetoelectronic instability}
%%%%%%%%%%%%%%%%%%%%%%%%%%%%%%%%%%%%%%%%%%%%%%%%%%
\label{sec::instability0}
%%%%%%%%%%%%%%%%%%%%%%%%%%%%%%%%%%%%%%%%%%%%%%%%%%

For completeness, we briefly outline the origin of the magnetoelectronic
instability for non-interacting electrons
$U=0$
(more details can be found in our
Ref.~\onlinecite{our_JETP_Lett}).
Imagine that the perfect ferromagnetic order in the substrate [described by
expression~(\ref{eq::M_FM})]
is distorted by weak canting deformation along $Oy$~axis
\begin{eqnarray}
%%%%%%%%%%%%%%%%%%%%%%%%%%%%%%%%%%%%%%%%%%%%%
\label{spatialmagn}
%%%%%%%%%%%%%%%%%%%%%%%%%%%%%%%%%%%%%%%%%%%%%
\mathbf{M} = M \left( 0,\, m,\, 1 + O( m ^2)
\right),
\end{eqnarray}
where 
$m = m(\mathbf{r}) = m (x, {\bf R})$,
$|m| \ll 1$,
represents the deviation of magnetization from axis $Oz$.

The canting increases the magnetic energy of the substrate
$E_{\rm m}$.
One can expect that weak deformation $m$ leads to quadratic correction to
the magnetic energy:
$\delta E_{\rm m} = O(m^2)$,
and
$\delta E_{\rm m} > 0$.
Due to non-zero $m$, the graphene electrons experience non-uniform Zeeman
field. As long as 
$h = \tau M m$
is small, it is possible to rely on the second-order perturbation theory to
evaluate the graphene-electron energy correction
$\delta E_{\rm e}$
caused by $m$. As it is always the case for second-order corrections, 
$\delta E_{\rm e}$
is non-positive
($\delta E_{\rm e} \leq 0$),
but in the limit of small $\tau$ and $M$ the total energy increases
($\delta E_{\rm m} + \delta E_{\rm e} > 0$),
implying the overall stability of the homogeneous ferromagnetic
configuration and the corresponding band structure.

Yet, there is a special type of canting $m$, for which this stability
argument does not work. We explained in
Ref.~\onlinecite{our_JETP_Lett}
that, because of the nested Fermi surface in our heterostructure, it is
possible to construct
$m = m ({\bf r})$
such that the corresponding correction 
$\delta E_{\rm e}$
is no longer perturbative, but rather non-analytical, and the total
correction
$\delta E_{\rm m} + \delta E_{\rm e} $
may become negative for suitable choice of parameters. This is the origin
of the magnetoelectronic instability.

To make our reasoning more concrete, let us consider $m$ such that
$m = m_\perp >0$
under carbon atoms belonging to sublattice $A$, and
$m = - m_\perp <0$
under atoms of sublattice $B$. In other words,
\begin{eqnarray}
%%%%%%%%%%%%%%%%%%%%%%%%%%%%%%%%%%%%%%%%%%%%
\label{eq::canted_condition}
%%%%%%%%%%%%%%%%%%%%%%%%%%%%%%%%%%%%%%%%%%%%
m (0, {\bf R}_A) = - m (0, {\bf R}_B) = m_\perp\,.
\end{eqnarray}
Schematically, the state with such a canting deformation is shown in
Fig.~\ref{fig::spec_numer_app}\,(b).
Deeper in the substrate and away from its surface, the homogeneous
magnetizations is restored [that is,
$m(x, {\bf R}) \rightarrow 0$
when
$x \rightarrow -\infty$].

Formally speaking, the canting deformations
satisfying
Eq.~(\ref{eq::canted_condition})
induce the hybridization between the electrons and holes at the Fermi
energy. Such a hybridization leads to the gap opening and non-analytical
contributions. Our calculations in this section will prove this fact
explicitly. As for the more general, symmetry-based, discussion of this
matter, interested readers may consult
Refs.~\onlinecite{sdw_inplane_magnet2007theor,ourAAbilayer2012prl}.

For the canting~(\ref{eq::canted_condition}),
the order-of-magnitude estimate for correction is, then,
%\begin{eqnarray}
$\delta E_{\rm m} \sim J M^2 m_\perp^2 {\cal N}_{\rm g}$,
%\end{eqnarray} 
where 
${\cal N}_{\rm g}$
is number of the unit cells in the graphene sample, and $J$ is the
ferromagnetic exchange constant. For a specific model of the ferromagnetic
substrate we derive (see
Appendix~\ref{app::weiss_model})
\begin{eqnarray}
\delta {\cal E}_{\rm m}
=
\frac{\pi}{3 \sqrt{3}} \frac{\zeta _\perp ^2 T_{\rm C}}{\zeta _0 ^2 t},
\end{eqnarray} 
where
$\delta {\cal E}_{\rm m} = \delta E_{\rm m}/({\cal N}_{\rm g} t)$
is the normalized magnetic energy per unit cell, and $\zeta_0$ and $\zeta_{\perp}$ are two dimensionless Zeeman fields
\begin{eqnarray}
    \zeta _0 = \frac{h}{t}, \quad
    \zeta _\perp = \frac{h_\perp}{t}\,.
\end{eqnarray}

To calculate the energy of the electrons we need to find the band structure
in the presence of the
canting~(\ref{eq::canted_condition}).
Since in this Section we neglect the electron-electron interaction
($U=0$)
the Hamiltonian for the graphene electrons may be expressed as in
Eq.~(\ref{eq::H_noninteract}),
with matrix
${\cal H}_{\bf q}$
being equal to
\begin{eqnarray}
%%%%%%%%%%%%%%%%%%%%%%%%%%%%%%%%%%%%%%%%%%%%%%%%%%
\label{eq::matrix_H}
%%%%%%%%%%%%%%%%%%%%%%%%%%%%%%%%%%%%%%%%%%%%%%%%%%
\hat{\cal H}_{\bf q}
=
\left( \begin{matrix}
    h & -t f_{\mathbf{q}} & -i h _{\perp}  & 0 \\
    -t f _{\mathbf{q}} ^{*} & h & 0 & i h _{\perp} \\
    i h _{\perp} & 0 & -h & -t f _{\mathbf{q}} \\
    0 & -i h _{\perp}  & -t f _{\mathbf{q}} ^{*} & -h
    \end{matrix} \right),
\end{eqnarray}
where
\begin{eqnarray} 
h_\perp = \tau M m_\perp.
\end{eqnarray} 
Note that 
$h_\perp$
preserves the translation symmetry of the hexagonal lattice, but it does
violate the symmetry between the two sublattices.

Diagonalizing 
${\cal H}_{\bf q}$
one finds the dispersion in the presence of the canting 
deformation~(\ref{eq::canted_condition})
\begin{eqnarray}
%%%%%%%%%%%%%%%%%%%%%%%%%%%%%%%%%%%%%%%%%%%%%%%%%% 
\label{eq::disp2}
%%%%%%%%%%%%%%%%%%%%%%%%%%%%%%%%%%%%%%%%%%%%%%%%%% 
\varepsilon^{(1, 2, 3, 4)}_\mathbf{q}
=
\pm \sqrt{h _{\perp} ^{2} + \left( h \pm t|f _\mathbf{q}| \right) ^2}\,.
\end{eqnarray}
The spectrum is plotted in
Fig.~\ref{fig::dispersion}\,(b).

The total electronic energy of graphene can be expressed as
$E_{\rm e} = t {\cal N}_{\rm g} {\cal E}_{\rm e}$,
where the total dimensionless energy 
${\cal E}_{\rm e}$
is a sum of two contributions coming from two bands
${\cal E}_{\rm e} = {\cal E}_{+} + {\cal E}_{-}$
defined as
\begin{eqnarray}
{\cal E}_{\pm}
=
- \int _0 ^3 d \zeta \, \rho (\zeta) \sqrt{\zeta_{\perp} ^2 +
    \left( \zeta _0 \pm \zeta \right) ^2}.
\end{eqnarray}
In this equation, the
dimensionless honeycomb lattice DOS
\begin{eqnarray}
%%%%%%%%%%%%%%%%%%%%%%%%%%%%%%%%%%%%%%%%%%%%%%%%%% 
\label{eq::statedensity}
%%%%%%%%%%%%%%%%%%%%%%%%%%%%%%%%%%%%%%%%%%%%%%%%%% 
\rho (\zeta)
=
\int \frac{d^2 \mathbf{k}}{V _{\rm BZ}}
	 \delta \left( \zeta - |f_{\mathbf{k}}| \right)
\end{eqnarray}
was introduced. For small 
$|\zeta|$
one can derive
\begin{eqnarray} 
%%%%%%%%%%%%%%%%%%%%%%%%%%%%%%%%%%%%%%%%%%%%%%%%%% 
\label{eq::statedensity_lin}
%%%%%%%%%%%%%%%%%%%%%%%%%%%%%%%%%%%%%%%%%%%%%%%%%% 
\rho (\zeta)
=
\frac{2 |\zeta|}{\sqrt{3} \pi} + \frac{2 |\zeta| ^{3/2}}{3 \pi}
+ \ldots,
\\
\nonumber 
\end{eqnarray} 
which is a generalization of the well-known expression for the linear 
asymptotic of the graphene DOS. 

Straightforward mathematical analysis of the expression for
${\cal E}_{-}$
reveals that, at small
$m_\perp$,
\begin{eqnarray}
{\cal E}_{-} (\zeta_\perp)
\approx
{\cal E}_{-} (0)
+
2 \rho(\zeta_0) \zeta_\perp^2 \ln \zeta_\perp + \ldots,
\end{eqnarray} 
where the ellipsis stands for the terms which are less singular. We see
that, due to the 
$m_\perp^2\ln m_\perp$
term, the total energy of the heterostructure always decreases when 
$m_\perp$
departs from zero, indicating the instability of the homogeneous
ferromagnetic state.

The equilibrium value of 
$m_\perp$
is the solution of the minimization condition
\begin{eqnarray}
%%%%%%%%%%%%%%%%%%%%%%%%%%%%%%%%%%%%%%%%%%%%%%%%%%
\label{eq::diff_gap0}
%%%%%%%%%%%%%%%%%%%%%%%%%%%%%%%%%%%%%%%%%%%%%%%%%% 
\frac{\partial} {\partial \zeta _\perp}
\left( {\cal E}_{\rm e} + \delta {\cal E}_{m} \right) = 0.
\end{eqnarray} 
It is solved in
Appendix~\ref{app::gap_derivation},
where we derive the following expression for the dimensionless field
$\zeta_\perp$
\begin{eqnarray} 
%%%%%%%%%%%%%%%%%%%%%%%%%%%%%%%%%%%%%%%%%%%%%%%%%%
\label{eq::gap_zeroU}
%%%%%%%%%%%%%%%%%%%%%%%%%%%%%%%%%%%%%%%%%%%%%%%%%% 
\zeta _\perp
=
\sqrt{12 \zeta _0}
\exp{\left[
	\frac{\sqrt{3} \pi}{4 \zeta _0}
	 \left( 
		\eta (\zeta_0)-\frac{2\pi T_{\rm C}}{3\sqrt{3}\zeta_0^2t} 
	\right) 
\right]}.
\end{eqnarray}
Function
$\eta (\zeta_0)$
in this expression is defined as
\begin{eqnarray}
%%%%%%%%%%%%%%%%%%%%%%%%%%%%%%%%%%%%%%%%%%%%%%%%%%
\label{eq::eta_def}
%%%%%%%%%%%%%%%%%%%%%%%%%%%%%%%%%%%%%%%%%%%%%%%%%% 
\eta (\zeta _0)
=
\int_{0}^3 
\frac{\left[ \rho (\zeta) - \rho (\zeta _0) \right] d \, \zeta}
{ \zeta - \zeta _0}
+
\int_{0}^3 \frac{ \rho (\zeta)  d \, \zeta} { \zeta + \zeta _0}.
\end{eqnarray}
At small 
$\zeta_0$
one can demonstrate (see
Appendix~\ref{app::eta_approx})
that
\begin{eqnarray} 
%%%%%%%%%%%%%%%%%%%%%%%%%%%%%%%%%%%%%%%%%%%%%%%%%%
\label{eq::eta_approx}
%%%%%%%%%%%%%%%%%%%%%%%%%%%%%%%%%%%%%%%%%%%%%%%%%% 
\eta (\zeta_0)
\approx
2I - \frac{2 \zeta_0}{\sqrt{3} \pi} \ln \left(\frac{3 }{\zeta_0}\right),
\end{eqnarray} 
where the constant
$I \approx 0.89$
is evaluated numerically. This approximation allows us to simplify
Eq.~(\ref{eq::gap_zeroU})
\begin{eqnarray}
\zeta_\perp
=
2 \zeta _0 \exp \left(
	\frac{C}{\zeta _0}
	-
	\frac{\pi^2 T_{\rm C}}{6 \zeta _0^3 t} 
\right).
\end{eqnarray} 
The coefficient $C$ in this formula is
$C \approx 2.42$.
The order parameter equals to
\begin{eqnarray}
%%%%%%%%%%%%%%%%%%%%%%%%%%%%%%%%%%%%%%%%%%%%%%%%%%
\label{eq::OP_simp}
%%%%%%%%%%%%%%%%%%%%%%%%%%%%%%%%%%%%%%%%%%%%%%%%%%
h_\perp = t \zeta_\perp = 
2 h \exp{\left( \frac{C t}{h}
	-
	\frac{\pi^2 T_{\rm C} t^2}{6 h^3} 
\right)},
\end{eqnarray} 
while the spectral gap is
$2 h_\perp$.
This relation for the gap can be compared against Eq.~(45) of
Ref.~\onlinecite{our_JETP_Lett}.
It is easy to see that, while the general structure of these two
expressions are almost identical, the gap value we found above always
exceeds the value found in
Ref.~\onlinecite{our_JETP_Lett}.
This is a consequence of the more advanced treatment of the higher-energy
single-electron states.

\section{Effects of the electron-electron interaction}
%%%%%%%%%%%%%%%%%%%%%%%%%%%%%%%%%%%%%%%%%%%%%%%%%%
\label{sec::interaction}
%%%%%%%%%%%%%%%%%%%%%%%%%%%%%%%%%%%%%%%%%%%%%%%%%%

In this section we discuss how electron-electron interaction influences the
magnetoelectronic instability. Specifically we will re-derive
Eq.~(\ref{eq::OP_simp})
in a model with
$U > 0$.
To reach this goal, we will use the mean field (MF) approximation. There are
several (perfectly equivalent) formulations to the MF approach. Below we
will use the variational version of the MF. To this end we introduce the MF
Hamiltonian
\begin{eqnarray}
%%%%%%%%%%%%%%%%%%%%%%%%%%%%%%%%%%%%%%%%%%%%%%%%%%
\label{eq::MF_Ham}
%%%%%%%%%%%%%%%%%%%%%%%%%%%%%%%%%%%%%%%%%%%%%%%%%% 
\hat{H}^{\rm MF}
=
\sum _{\mathbf{q}}
    \hat\Phi_{\bf q}^\dag
    \hat{\cal H}_{\bf q}^{\rm MF}
    \hat\Phi_{\bf q}^{\vphantom{\dag}},
\\
\hat{\cal H}_{\bf q}^{\rm MF}
=
t \left( \begin{matrix}
    \Tilde \zeta_0 & -f_{\mathbf{q}} & -i \Tilde \zeta_{\perp}  & 0 \\
    -f _{\mathbf{q}} ^{*} & \Tilde \zeta_0 & 0 & i \Tilde \zeta _{\perp} \\
    i \Tilde \zeta _{\perp} & 0 & -\Tilde \zeta_0 & -f _{\mathbf{q}} \\
    0 & -i \Tilde \zeta _{\perp}  & -f _{\mathbf{q}} ^{*} & -\Tilde \zeta_0
    \end{matrix} \right).
\end{eqnarray}
The ground state 
$\left|\Psi_{\rm MF} \right>$
%=\left|\Psi_{\rm MF} (\Tilde \zeta_0, \Tilde \zeta_\perp)\right>$
of
$\hat{H}^{\rm MF}$
acts as our variational wave function. Dimensionless quantities
$\Tilde \zeta_0$
and
$\Tilde \zeta_\perp$
will serve as the optimization parameters for our variational ansatz.
Matrix
$\hat{\cal H}_{\bf q}^{\rm MF}$
has the same structure as
$\hat{\cal H}_{\bf q}$
in
Eq.~(\ref{eq::matrix_H}).
In other words, we assume that the interaction renormalizes parameters
$\zeta_0$
and
$\zeta_\perp$
responsible for the magnetoelectronic instability. As we will see below,
this renormalization accounts for the interaction-driven enhancement of the
instability.

Using the symbol
$\langle \ldots \rangle$
to denote matrix element with respect to
$\left| \Psi_{\rm MF} \right>$,
we can express the total dimensionless variational energy as
\begin{eqnarray}
%%%%%%%%%%%%%%%%%%%%%%%%%%%%%%%%%%%%%%%%%%%%%%%%%%
\label{eq::E_var}
%%%%%%%%%%%%%%%%%%%%%%%%%%%%%%%%%%%%%%%%%%%%%%%%%% 
{\cal E} ^{\rm var}
=
({\cal N}_{\rm g} t)^{-1} \langle \hat{H} \rangle  + \delta {\cal E}_{\rm m}.
\end{eqnarray}
Adjusting
$\zeta_\perp$,
$\Tilde \zeta_0$,
and
$\Tilde \zeta_\perp$,
the minimum of
${\cal E}^{\rm var}$
must be found. Differentiating
${\cal E}^{\rm var}$
with respect to
$\Tilde \zeta_\perp$,
we obtain
\begin{eqnarray}
%%%%%%%%%%%%%%%%%%%%%%%%%%%%%%%%%%%%%%%%%%%%%%%%%% 
\label{eq::perpdiff}
%%%%%%%%%%%%%%%%%%%%%%%%%%%%%%%%%%%%%%%%%%%%%%%%%% 
%\frac{\partial \langle \hat{H} \rangle}{\partial \Tilde{\zeta} _{\perp}} = 
2 t \left( \zeta _\perp - \Tilde{\zeta} _{\perp} \right ) - 
U \langle S ^{y} \rangle = 0,
%\frac{\partial \langle S ^{y} \rangle }{\partial \Tilde{\zeta} _\perp} = 0,
\end{eqnarray}
where
$\langle S ^{y} \rangle$
is the average magnetization projection on a site belonging to the
sublattice $A$, that is
$\langle S ^{y}_{{\bf R}_A} \rangle = \langle S^{y} \rangle$.
Using the Hellmann\,-Feynman theorem (see
Appendix~\ref{app::MF_derivation}),
we derive
\begin{eqnarray}
%%%%%%%%%%%%%%%%%%%%%%%%%%%%%%%%%%%%%%%%%%%%%%%%%% 
\label{eq::cantaxisy}
%%%%%%%%%%%%%%%%%%%%%%%%%%%%%%%%%%%%%%%%%%%%%%%%%% 
\langle S ^{y} \rangle =
-\frac{\Tilde{\zeta} _\perp}{2}
\left[ 
	\eta (\zeta_0) + 
	\frac{2 \Tilde{\zeta} _0}{\sqrt{3} \pi} 
	\ln{\left(\frac{12\Tilde\zeta_0}{\Tilde\zeta_\perp^2} \right)} 
\right].
\end{eqnarray}
Note that
$\langle S ^{y}_{{\bf R}_B} \rangle = - \langle S^{y} \rangle$.
In other words, the average magnetization $y$-projections have different
signs on different sublattices.
(For $z$ axis one has
$\langle S ^{z}_{{\bf R}_A} \rangle = 
 \langle S ^{z}_{{\bf R}_B} \rangle = \langle S^{z} \rangle$.)

Differentiation over
$\Tilde{\zeta} _0$
allows us to obtain the second mean field equation 
\begin{eqnarray}
%%%%%%%%%%%%%%%%%%%%%%%%%%%%%%%%%%%%%%%%%%%%%%%%%% 
\label{eq::parrdiff}
%%%%%%%%%%%%%%%%%%%%%%%%%%%%%%%%%%%%%%%%%%%%%%%%%% 
%\frac{\partial \langle \hat{H} \rangle}{\partial \Tilde{\zeta} _0} =
%\left[ 
2 t \left( \zeta _0 - \Tilde{\zeta} _0 \right)
- U \langle S ^{z} \rangle = 0.
%\right] \frac{\partial \langle S ^{z} \rangle }{\partial \Tilde{\zeta}} = 0
\end{eqnarray}
As above, to calculate
$\langle S ^{z} \rangle$,
one can invoke the Hellmann\,-Feynman theorem and find
\begin{eqnarray}
%%%%%%%%%%%%%%%%%%%%%%%%%%%%%%%%%%%%%%%%%%%%%%%%%% 
\label{eq::cantaxisz}
%%%%%%%%%%%%%%%%%%%%%%%%%%%%%%%%%%%%%%%%%%%%%%%%%% 
\langle S^{z} \rangle
%= \int _0 ^3 \frac{d \zeta \, \rho (\zeta) 
%    (\zeta - \Tilde{\zeta} _0)}{\sqrt{\Tilde{\zeta} _{\perp} ^2 +
%    \left( \Tilde{\zeta} _0 - \zeta \right) ^2}} -
%    \\
%    \nonumber
%    \int _0 ^3 \frac{d \zeta \, \rho (\zeta) 
%    (\zeta + \Tilde{\zeta} _0)}{\sqrt{\Tilde{\zeta} _{\perp} ^2 +
%    \left( \Tilde{\zeta} _0 + \zeta \right) ^2}} \approx
%    \\
%    \nonumber
%    \approx
%    \int _0 ^3 \frac{d \zeta \, \rho (\zeta) 
%    ( \zeta - \Tilde{\zeta _0} ) }{\left| \Tilde{\zeta} _0 - \zeta \right|} -
%    \int _0 ^3 \frac{d \zeta \, \rho (\zeta) 
%    ( \zeta + \Tilde{\zeta _0})}{\left| \Tilde{\zeta} _0 + \zeta \right|}
%    =
%    \\
%    \nonumber
%    = \int _0 ^3 d \zeta \, \rho (\zeta) \left[ {\rm sgn} (\zeta - 
%    \Tilde{\zeta} _0) - 1 \right] =
%    \\
%    \nonumber
%    = -\int _0 ^{\Tilde{\zeta} _0 }  d \zeta \, \rho (\zeta) \simeq 
\approx    - \frac{\sqrt{3}}{\pi} \Tilde{\zeta} _0 ^2   .
\end{eqnarray}
Deriving this relation, we used 
Eq.~(\ref{eq::statedensity_lin})
valid for small
$\Tilde{\zeta}_0$.
Since
$\langle S^{z} \rangle = O(\Tilde\zeta_0^2)$,
the last term in
Eq.~(\ref{eq::parrdiff})
can be neglected, and we conclude
\begin{eqnarray}
    \Tilde{\zeta}_0 \approx \zeta _0.
\end{eqnarray}
This formula implies that the interaction does not introduce significant
renormalization to the homogeneous Zeeman field induced by the substrate.

\begin{figure}
    \centering
    \includegraphics[width=0.85\columnwidth]{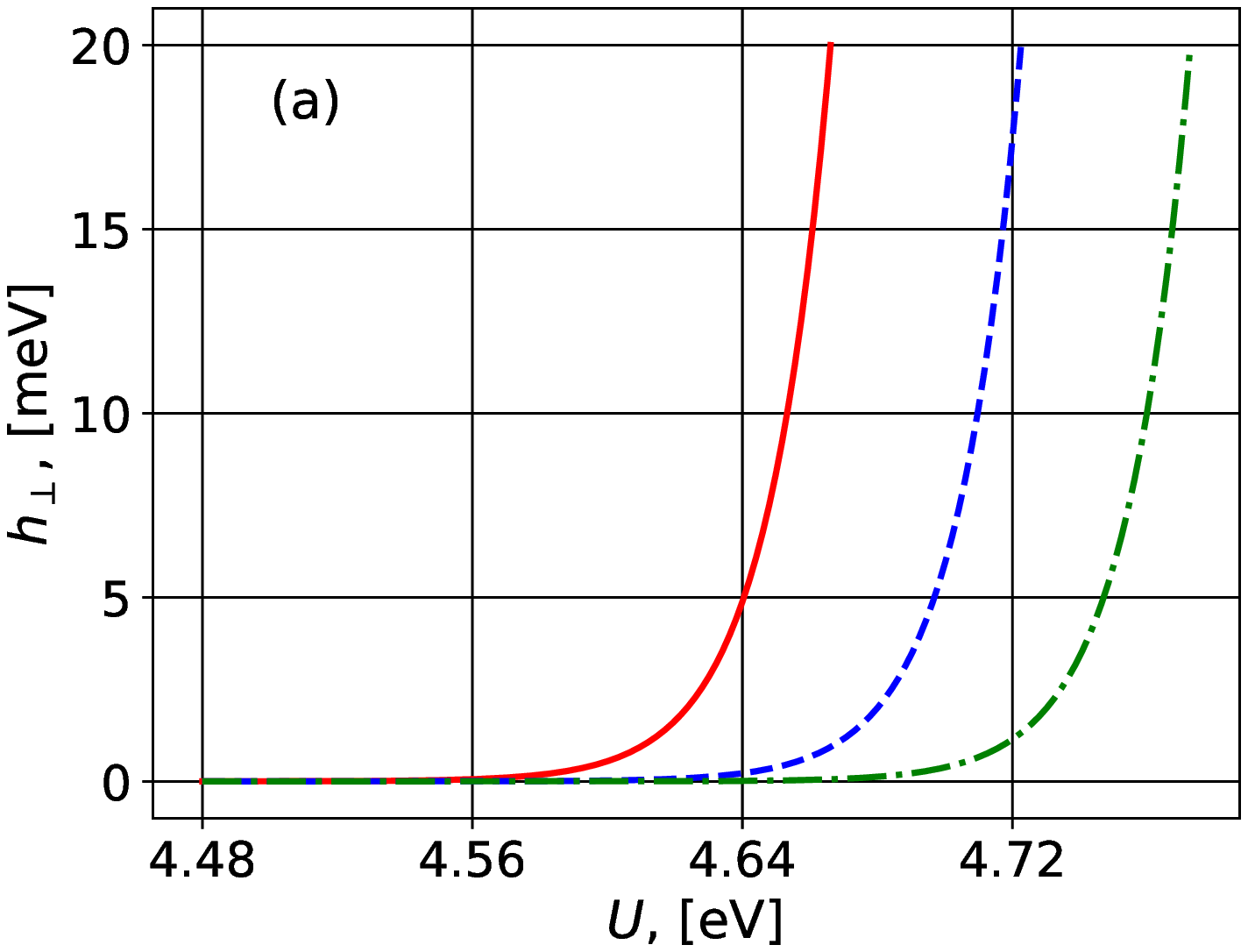}
    \includegraphics[width=0.85\columnwidth]{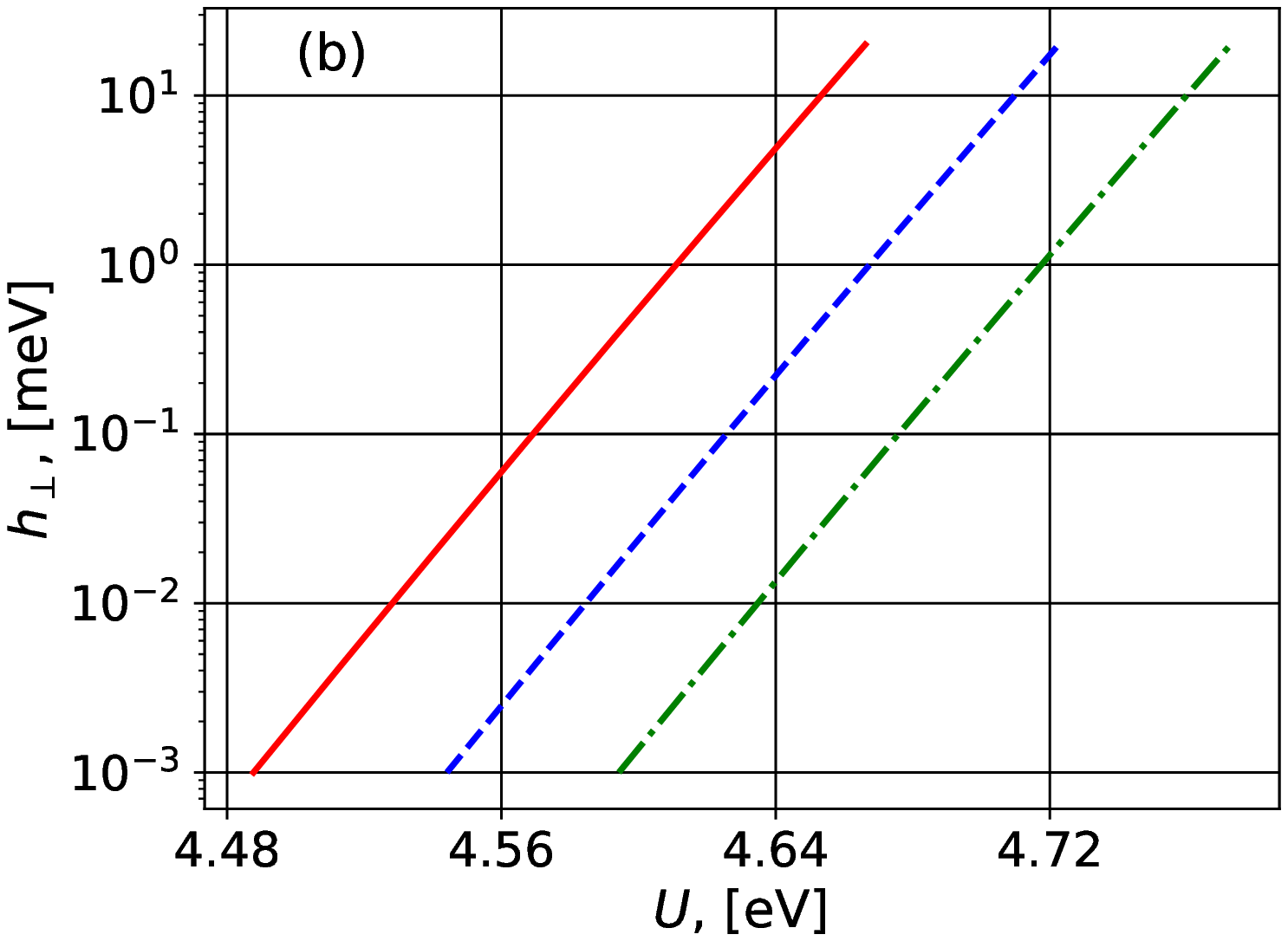}
    \caption{Order parameter $h_\perp$ as function of the interaction
parameter $U$, for various
$T_{\rm C}$,
on linear (a) and log (b) scale. 
Solid (red) curve for 
$T_{\rm C} = 16$\,K,
dashed (blue) curve for 
$T_{\rm C} = 17$\,K,
dashed-dotted (green) curve for 
$T_{\rm C} = 18$\,K.
For these curves we use
$h = 20$\,meV
(or, equivalently,
$h = 300$\,T).
%%%%%%%%%%%%%%%%%%%%%%%%%%%%%%%%%%%%%%%%
\label{fig::Delta}
%%%%%%%%%%%%%%%%%%%%%%%%%%%%%%%%%%%%%%%%
    }
\end{figure}

Finally, we minimize
${\cal E}^{\rm var}$
with respect to
$\zeta _{\perp}$.
We obtain
\begin{eqnarray}
%%%%%%%%%%%%%%%%%%%%%%%%%%%%%%%%%%%%%%%%%%%%%%%%%%
\label{eq::var_zeta0}
%%%%%%%%%%%%%%%%%%%%%%%%%%%%%%%%%%%%%%%%%%%%%%%%%% 
\frac{\partial {\cal E} ^{\rm var}}{\partial \zeta _\perp}
=
2 \langle S^y \rangle
+
\frac{2 \pi}{3 \sqrt{3}} 
		\frac{T _{\rm C}}{t \zeta _0 ^2} \zeta _\perp = 0.
\end{eqnarray}
Collecting
Eqs.~(\ref{eq::var_zeta0}),
(\ref{eq::parrdiff})
and~(\ref{eq::cantaxisy}),
we obtain the following system of equation 
\begin{eqnarray}
    \begin{cases}
         \zeta _\perp - \Tilde{\zeta} _\perp
		= \frac{U}{2 t} \langle S^y \rangle,
        \\
        - 2 \langle S^y \rangle 
		= \frac{2 \pi}{3 \sqrt{3}} 
		\frac{T _{\rm C}}{t \zeta _0 ^2} \zeta _\perp
        \\
        - 2 \langle S^y \rangle
		 = \Tilde{\zeta}_\perp 
		\left(\eta (\zeta _0) +
	    		\frac{2 \zeta _0}{\sqrt{3} \pi} 
			\ln{\left| \frac{12 \zeta_0}
				{\Tilde{\zeta} _\perp ^2} \right|} 
		\right),
    \end{cases}
\end{eqnarray}
where the last equation can be simplified with the help of
Eq.~(\ref{eq::eta_approx})
Solving this system, one determines 
$\Tilde{\zeta} _\perp$
and finds the order parameter
$h_\perp = t \Tilde{\zeta} _\perp$
\begin{eqnarray}
%%%%%%%%%%%%%%%%%%%%%%%%%%%%%%%%%%%%%%%%%%%%%%%%%%
\label{eq::gap_finU}
%%%%%%%%%%%%%%%%%%%%%%%%%%%%%%%%%%%%%%%%%%%%%%%%%% 
h_\perp
= 
2 h \exp \left( 
		\frac{C t}{h}
		-
		\frac{\sqrt{3} \pi^2 T_{\rm C}t^2}
			{6\sqrt{3}h^3 + \pi UT_{\rm C} h}
	 \right).
\quad
%\end{multline}
\end{eqnarray}
This expression coincide with
Eq.~(\ref{eq::gap_zeroU})
in the limit of vanishing $U$, as expected. The gap increases when $U$
grows. The resultant values of $h_\perp$ versus $U$ are plotted in
Fig.~\ref{fig::Delta}
for various parameters choices.

\section{Discussion}
%%%%%%%%%%%%%%%%%%%%%%%%%%%%%%%%%%%%%%%%%%%%%%%%%%
\label{sec::discussion}
%%%%%%%%%%%%%%%%%%%%%%%%%%%%%%%%%%%%%%%%%%%%%%%%%%

In the previous sections we demonstrated that graphene on a ferromagnetic
substrate is susceptible to cooperative magnetoelectronic instability. Near
the interface between the substrate and the graphene the instability
generates finite canting deformation of the ferromagnetic polarization.
The canting, in turn, leads to the spectral gap in the electronic spectrum
of the graphene. Our analysis generalizes the study of
Ref.~\onlinecite{our_JETP_Lett}
to include the effects of finite electron-electron interaction. As 
Eq.~(\ref{eq::gap_finU})
proves, the interaction enhances the instability.

The instability can be potentially detected in a transport experiment
measuring the temperature dependence of the graphene conductivity
$\sigma = \sigma (T)$.
The instability will manifest itself as a sharp decrease of 
$\sigma (T)$
when $T$ drops below the characteristic energy scale
$\sim h_{\perp}$. 
This implies that, for the instability to be observable in experiment, the
value of $h_{\perp}$ must be sufficiently large. To assess $h_{\perp}$, one can
examine 
Fig.~\ref{fig::Delta},
which shows the plots of $h_{\perp}$ versus $U$ for different values of
$T_{\rm C}$.
The graphs in the latter figure reveals that, while
at
$U = 0$
the order parameter might be very weak, a realistic electron-electron interaction 
drastically, by orders of magnitude, enhances the equilibrium
value of
$h_{\perp}$,
making it sufficiently strong to be detected in an experiment. Specifically, for the
superstructure of graphene on EuS substrate, described in
Ref.~\onlinecite{Wei2016},
one has
$T_{\rm C} = 16$\,K
and
$h = 300$\,T
(equivalently,
$h = 20$\,meV).
For such parameters
$h_{\perp} \gtrsim 10$\,meV
for
$U \gtrsim 4.65$\,eV.
Examining 
Eq.~(\ref{eq::gap_finU})
for $h_{\perp}$ we notice that, as far as the interaction strength $U$ is
concerned, two regimes can be identified. To illustrate this, let us
re-write
Eq.~(\ref{eq::gap_finU})
as follows
\begin{eqnarray}
%%%%%%%%%%%%%%%%%%%%%%%%%%%%%%%%%%%%%%%%%%%%%%%%%%
\label{eq::gap_Ux}
%%%%%%%%%%%%%%%%%%%%%%%%%%%%%%%%%%%%%%%%%%%%%%%%%% 
h_{\perp} 
= 
2 h \exp \left( 
		\frac{C t}{h}
		-
		\frac{\sqrt{3} \pi t^2/h} {U_* + U}
	 \right),
\end{eqnarray}
where the characteristic interaction strength is
\begin{eqnarray}
U_* = \frac{6\sqrt{3} h^2}{\pi T_{\rm C}}.
\end{eqnarray}
We see that, when
$U \ll U_*$,
one can neglect $U$ relative to 
$U_*$
in
Eq.~(\ref{eq::gap_Ux}).
In this limit, the effects of the interaction are weak, and the instability
is driven exclusively by the cooperation between the substrate magnetic
subsystem and the electrons of the graphene. This regime is studied in
Ref.~\onlinecite{our_JETP_Lett}.

In the opposite case
($U \gg U_*$)
one can treat
$U_*$
as a small correction to $U$, neglecting
$U_*$
in the zeroth-order approximation. If 
$U_*$
is removed from
Eq.~(\ref{eq::gap_Ux}),
then
$T_{\rm C}$
does not enter the expression for $h_{\perp}$, indicating that the cooperation
between the canting deformation and the electrons is no longer important.
Instead, the only remaining role of the ferromagnetic substrate is to
generate a Fermi surface with finite DOS
$\rho \propto h$
at the Fermi energy. In this regime, the nested Fermi surface is driven
into the SDW-like insulating state by the electron-electron interaction, as
discussed in
Ref.~\onlinecite{sdw_inplane_magnet2007theor}.

The two regimes are connected by a crossover which occurs at
$U \sim U_*$.
In the crossover region, both electron-electron interaction effects and the
canting deformation in the substrate cooperate together equally to induce
the instability.

To conclude, we studied the influence of the electron-electron interaction
on the magnetoelectronic instability in graphene on the ferromagnetic
substrate. We demonstrated that the interaction enhances the instability
significantly: even a moderate strength interaction can increase the
characteristic energy scale by several orders of magnitude. Our findings
suggest that the instability could be detected experimentally in realistic
settings.

\begin{acknowledgments}
The research is funded by the Russian Science Foundation grant
No.~22-22-00464,
\url{https://rscf.ru/project/22-22-00464/}.
\end{acknowledgments}

\appendix

\section{Magnetic energy correction}
%%%%%%%%%%%%%%%%%%%%%%%%%%%%%%%%%%%%%%%%%%%%%%%%%%
\label{app::weiss_model}
%%%%%%%%%%%%%%%%%%%%%%%%%%%%%%%%%%%%%%%%%%%%%%%%%%

In this appendix we calculate the magnetic energy of the substrate with and
without the canting deformation of the ferromagnetic magnetization. Our
starting point is the Weiss model of a ferromagnet. In the Weiss
approximation, the magnetic energy of the substrate is
\begin{eqnarray}
\label{Kuri-Veis}
E_m
=
- \frac{J}{2} \sum_{i \boldsymbol{\delta}}
    \mathbf{M}_{{\bf r}_i} \mathbf{M}_{{\bf r}_i + \boldsymbol{\delta}}\,.
\end{eqnarray}
Here $J$ is the exchange integral,
${\bf r}_i$
denotes position of $i$'th ``atom" in three-dimensional lattice of the
substrate, vectors
$\boldsymbol{\delta}$
link a given atom with its nearest neighbors, and
${\bf M}_{{\bf r}_i}$
is the magnetization on the $i$'th atom. We assume that vectors
${\bf M}_{{\bf r}_i}$
have the same length:
$|{\bf M}_{{\bf r}_i}| = M$
for any
$i$.

For a model of this type, we can derive expression for Curie temperature:
\begin{eqnarray}
%%%%%%%%%%%%%%%%%%%%%%%%%%%%%%%%%%%%%%%%%%%%%%%%%%
\label{eq::Curie_temperature}
%%%%%%%%%%%%%%%%%%%%%%%%%%%%%%%%%%%%%%%%%%%%%%%%%%
T_{\rm C} = \frac{J M^2 Z}{3},
\end{eqnarray}
where $Z$ is the number of the nearest neighbors of an atom. This
expression can be obtained using the mean-field method applied to
energy~(\ref{Kuri-Veis}).
The advantage of
Eq.~(\ref{eq::Curie_temperature})
stems from the fact that it allows us to estimate $J$, using
often-available experimental data for the Curie temperature,
low-temperature substrate magnetization $M$, and $Z$.

Denoting the number of atoms in the substrate as
${\cal N}$,
we can express the energy of the perfectly ordered ferromagnetic
configuration as
\begin{eqnarray} 
E_\| = - \frac{\cal N}{2} Z J  M ^2\,,
\end{eqnarray} 
which is valid at
$T \ll T_{\rm C}$.
For configurations with weak and smooth deviations from the homogeneously
ordered state the energy becomes
$E_m \approx E_\| + \delta E_{\rm m}$,
where the correction
$\delta E_{\rm m}$
equals to
\begin{eqnarray}
%%%%%%%%%%%%%%%%%%%%%%%%%%%%%%%%%%%%%%%%%%%%%%%%%%
\label{eq::magnetic_correction}
%%%%%%%%%%%%%%%%%%%%%%%%%%%%%%%%%%%%%%%%%%%%%%%%%%
\delta E_m 
\approx
\frac{J M ^2}{2 a _0} \int_V\!\! (\nabla m )^2 dV\,,
\end{eqnarray}
where we assumed that substrate material has cubic lattice
($Z = 6$)
whose lattice constant is
$a_0$
(generalizations beyond these two assumptions are trivial).
The dimensionless cant
$m = m({\bf r})$,
$|m| \ll 1$,
is introduced by
Eq.~(\ref{spatialmagn}).
It is treated here as a function of a three-dimensional continuous variable
${\bf r} = (x, {\bf R})$,
while integration in
Eq.~(\ref{eq::magnetic_correction})
is performed over the volume of the substrate 
($x < 0$).

We now want to apply
Eq.~(\ref{eq::magnetic_correction})
for the evaluation of the magnetic energy increase
$\delta E_{\rm m}$
caused by the canting deformation shown in
Fig.~\ref{fig::spec_numer_app}\,(b).
Our goal is (i)~to find $m$ which satisfies boundary
condition~(\ref{eq::canted_condition})
at
$x = 0$,
while (ii)~delivering the minimum to the
functional~(\ref{eq::magnetic_correction}).

However, the boundary 
condition~(\ref{eq::canted_condition})
was formulated for a function on a lattice. For a continuous
approximation that we employ in this Appendix, such a boundary condition is
incomplete. Indeed, it fixes the values of $m$ on a discrete set of points
only, thus, on
$\mathbb{R}^2$,
there are infinitely many non-identical functions satisfying
Eq.~(\ref{eq::canted_condition}).
To mend this problem we extend the boundary condition
\begin{eqnarray}
%%%%%%%%%%%%%%%%%%%%%%%%%%%%%%%%%%%%%%%%%%%%%
\label{eq::bounadry_conditions}
%%%%%%%%%%%%%%%%%%%%%%%%%%%%%%%%%%%%%%%%%%%%%
m (0, {\bf R})  =
- \frac{2}{3 \sqrt{3}} m _\perp 
\sum _{\mathbf{K}} \sin{(\mathbf{K} \mathbf{R})}\,,
\quad {\bf R} \in \mathbb{R}^2.
\end{eqnarray}
where vector
$\mathbf{K}$
runs over the following list of values
\begin{eqnarray}
%%%%%%%%%%%%%%%%%%%%%%%%%%%%%%%%%%%%%%%%%%%%%
\label{vecsys}
%%%%%%%%%%%%%%%%%%%%%%%%%%%%%%%%%%%%%%%%%%%%%
\mathbf{K} = \mathbf{b}_1, \quad
\mathbf{K} = \mathbf{b}_2, \quad
\mathbf{K} = - \mathbf{b}_1 - \mathbf{b}_2.
\end{eqnarray}
The advantage of this sum is that not only it is compatible with the
Eq.~(\ref{eq::canted_condition}),
but it contains the minimum number (six) of plane waves, while all spatial
frequencies
$|{\bf K}| = 4\pi/(3 a_0)$
have the lowest possible values compatible with
Eq.~(\ref{eq::canted_condition}).
Adding more plane waves with shorter wave lengths leads to more positive
contributions to the functional
$\delta E_{\rm m}$.

To proceed with (i) and (ii) formulated above, we derive the Laplace
equation 
\begin{eqnarray}
\nabla^2 m  = 0\,,
\end{eqnarray}
for $m$ within the substrate, with 
Eq.~(\ref{eq::bounadry_conditions})
at the interface. The solution to this mathematical problem is 
\begin{eqnarray}
m(\mathbf{r})
=
-\frac{2}{3 \sqrt{3}} m_\perp \exp{\left(  |\mathbf{K}| x \right)}
	\sum _{\mathbf{K}} \sin{(\mathbf{K} \mathbf{R})}\,.
\end{eqnarray}
This expression demonstrates that, as expected, the canting deformation of
the magnetization is the strongest directly at the surface
($x=0$),
but it quickly decreases deep into the substrate. Substituting this 
$m=m({\bf r})$
into
Eq.~(\ref{eq::magnetic_correction}),
we obtain 
\begin{eqnarray}
%%%%%%%%%%%%%%%%%%%%%%%%%%%%%%%%%%%%%%%%%%%%%%%%%% 
\label{freecorrection}
%%%%%%%%%%%%%%%%%%%%%%%%%%%%%%%%%%%%%%%%%%%%%%%%%% 
\delta E _{m} = \frac{4 \pi}{27} \frac{J m _\perp ^2 M ^2 {\cal S}}{a _0 ^2} =
\frac{4 \pi}{27} \frac{J h _\perp ^2 {\cal S}}{a_0^2 \tau^2}\,.
\end{eqnarray}
Alternatively, we can write using dimensionless quantities that
\begin{eqnarray}
%%%%%%%%%%%%%%%%%%%%%%%%%%%%%%%%%%%%%%%%%%%%%%%%%% 
\label{eq::dimlesscorrection}
%%%%%%%%%%%%%%%%%%%%%%%%%%%%%%%%%%%%%%%%%%%%%%%%%% 
\delta {\cal E}_{m}
=
\frac{\pi}{3 \sqrt{3}} \frac{\zeta _\perp ^2 T_{\rm C}}{\zeta _0 ^2 t}.
\end{eqnarray}
As we mentioned in
Sec.~\ref{sec::instability0},
the magnetic energy associated with the cant is quadratic in
$m_ \perp$
and does not contain any non-analytic contributions.

\section{Single-particle gap value}
%%%%%%%%%%%%%%%%%%%%%%%%%%%%%%%%%%%%%%%%%%%%%%%%%%
\label{app::gap_derivation}
%%%%%%%%%%%%%%%%%%%%%%%%%%%%%%%%%%%%%%%%%%%%%%%%%%

In this Appendix we provide detailed derivation of 
Eq.~(\ref{eq::gap_zeroU})
for the dimensionless order parameter 
$\zeta_\perp$.
The calculations presented below improves the derivation of
Eq.~(30)
in
Ref.~\onlinecite{our_AA_prb2013}.
Our starting point is
Eq.~(\ref{eq::diff_gap0})
which we expand as follows
\begin{widetext}
\begin{eqnarray}
%%%%%%%%%%%%%%%%%%%%%%%%%%%%%%%%%%%%%%%%%%%%%%%%%% 
\label{eq::diff_gap}
%%%%%%%%%%%%%%%%%%%%%%%%%%%%%%%%%%%%%%%%%%%%%%%%%% 
    \frac{\partial \left( {\cal E}_{1} + {\cal E}_{2} + 
    \delta {\cal E}_{m} \right)}{\partial \zeta _\perp}
    =
    \zeta_\perp \left( -\int_{0}^3 \frac{\rho (\zeta) d \, \zeta}
    {\sqrt{\zeta _\perp ^2 + \left( \zeta - \zeta _0 \right)^{2}}} 
    -
    \int_{0}^3 \frac{\rho (\zeta) d \, \zeta}
    {\sqrt{\zeta _\perp ^2 + \left( \zeta + \zeta _0 \right)^{2}}}  
    +
    \frac{2 \pi}{3 \sqrt{3}} \frac{T_{\rm C}}{\zeta _0 ^2 t}
 \right)
\approx 
    \\
    \nonumber
    \approx 
    \zeta_\perp \left( 
        \frac{2 \pi}{3 \sqrt{3}} \frac{T_{\rm C}}{\zeta _0 ^2 t} 
%    \int_{0}^3 
%	\frac{\left[ \rho (\zeta) - \rho (\zeta _0) \right] d \, \zeta}
%    {| \zeta - \zeta _0|} 
%    - \int_{0}^3 \frac{ \rho (\zeta) d \, \zeta}
%    {| \zeta + \zeta _0|} -
	- \rho (\zeta_0) \int_{0}^3 \frac{ d \, \zeta}
    {\sqrt{\zeta _\perp ^2 + \left( \zeta - \zeta _0 \right)^{2}}}
 	- \eta (\zeta_0) 
    \right) = 0,
\end{eqnarray}
\end{widetext}
where function
$\eta (\zeta_0)$
is defined by
Eq.~(\ref{eq::eta_def}).
The integral
$\rho (\zeta_0) \int_{0}^3 { d \, \zeta}
(\zeta _\perp ^2 + \left( \zeta - \zeta _0 \right)^{2})^{-1/2}$
in~(\ref{eq::diff_gap})
can be found explicitly
\begin{eqnarray}
\rho (\zeta_0) \int_{0}^3 \frac{ d \, \zeta}
{\sqrt{\zeta _\perp ^2 + \left( \zeta - \zeta _0 \right)^{2}}}
\approx \frac{2 \zeta_0}{\sqrt{3} \pi} 
\ln{\left( \frac{12 \zeta _0}{\zeta _\perp ^2} \right)}.
\end{eqnarray} 
As a result we can solve
equation~(\ref{eq::diff_gap}) 
%\begin{eqnarray}
%    \frac{6 \sqrt{3} \pi}{27} \frac{T_{\rm C}}{\zeta _0 ^2 t} - 2 I = \frac{4 |\zeta _0|}{\pi \sqrt{3}}
%    \ln{\left| \frac{12 \zeta _0}{\zeta _\perp ^2} \right| }.
%\end{eqnarray}
and obtain value of
$\zeta _\perp$
\begin{eqnarray}
    \zeta _\perp
=
\sqrt{12 \zeta _0}
\exp{\left\{ \frac{\sqrt{3} \pi}{4 \zeta _0} 
	\left( 
		\eta(\zeta_0) 
		-
		\frac{2 \pi T_{\rm C}}{3 \sqrt{3} \zeta _0 ^2 t}
	\right)
\right\}},
\end{eqnarray}
which is
Eq.~(\ref{eq::gap_zeroU}).

\section{Approximation for $\eta (\zeta_0)$}
%%%%%%%%%%%%%%%%%%%%%%%%%%%%%%%%%%%%%%%%%%%%%%%%% 
\label{app::eta_approx}
%%%%%%%%%%%%%%%%%%%%%%%%%%%%%%%%%%%%%%%%%%%%%%%%%% 

In this Appendix we derive the approximate form for
$\eta (\zeta_0)$
valid in the limit of small
$\zeta_0$.
We start by noting that the low-$\zeta_0$ singularities of the integral
$\int_{0}^3 d \, \zeta
{\left[ \rho (\zeta) - \rho (\zeta _0) \right] }/ {| \zeta - \zeta _0|}$
are weak, and it is sufficient to to approximate the integral by a constant
\begin{eqnarray}
%%%%%%%%%%%%%%%%%%%%%%%%%%%%%%%%%%%%%%%%%%%%%%%%%%
\label{eq::less_singular}
%%%%%%%%%%%%%%%%%%%%%%%%%%%%%%%%%%%%%%%%%%%%%%%%%% 
    \int_{0}^3 \frac
	{\left[ \rho (\zeta) - \rho (\zeta _0) \right] d \, \zeta}
	{| \zeta - \zeta _0|} = I + O(\zeta_0),
	\\
	I = \int _0 ^3 \frac{\rho (\zeta) d \zeta}{\zeta} \approx 0.89.
\end{eqnarray}
To prove that the ignored terms in 
Eq.~(\ref{eq::less_singular})
are indeed
$O(\zeta_0)$,
we split the integration interval into two sub-intervals
$\int_0^3 \ldots = \int_0^{\zeta^*} \ldots + 
\int_{\zeta^*}^3 \ldots$,
where 
$\zeta^*$
satisfies
$\zeta_0 < \zeta^* \ll 1$.
One can write
\begin{eqnarray}
\int_{\zeta^*}^3 \frac
	{\left[ \rho (\zeta) - \rho (\zeta _0) \right] d \, \zeta}
	{| \zeta - \zeta _0|}
= 
\int_{\zeta^*}^3 \frac {\rho (\zeta) d \, \zeta}{\zeta}
-
\\
\nonumber 
\rho (\zeta_0) \int_{\zeta^*}^3 \frac {d \, \zeta}{\zeta - \zeta_0}
+
\zeta_0 \int_{\zeta^*}^3 
	\frac{\rho (\zeta) d \, \zeta}{\zeta (\zeta - \zeta_0)}.
\end{eqnarray} 
In the right-hand side of this relation, the integrands are non-singular
for
$\zeta^* < \zeta < 3$.
The first integral is independent of 
$\zeta_0$,
while two other explicitly belong to
$O(\zeta_0)$
class.

As for the integral from 0 to
$\zeta^*$,
due to smallness of
$\zeta^*$,
the
expansion~(\ref{eq::statedensity_lin})
can be used. Thus
\begin{eqnarray}
%%%%%%%%%%%%%%%%%%%%%%%%%%%%%%%%%%%%%%%%%%%%%%%%%%
\label{eq::ratio_approx}
%%%%%%%%%%%%%%%%%%%%%%%%%%%%%%%%%%%%%%%%%%%%%%%%%%
\frac{\rho (\zeta) - \rho (\zeta _0) }{\zeta - \zeta _0} 
\approx
	\frac{2}{\sqrt{3} \pi } 
	+
	\frac{2 \sqrt{\zeta}}{3\pi}
	+
	\frac{2 \zeta_0}{3\pi (\sqrt{\zeta_0} + \sqrt{\zeta})}.
\end{eqnarray} 
Deriving this representation we used the following relation
%\begin{eqnarray}
$\frac{x^{3/2} - y^{3/2}}{x - y}
=
\frac{x + \sqrt{xy} + y}{x^{1/2} + y^{1/2}}.
$%\end{eqnarray} 
We see from
Eq.~(\ref{eq::ratio_approx})
that the integral over the sub-interval
$(0, \zeta^*)$
equals to a $\zeta_0$-independent constant plus
$O(\zeta_0)$
corrections, as
Eq.~(\ref{eq::less_singular})
implies.

The second integral in
Eq.~(\ref{eq::eta_def})
can be transformed as follows
\begin{eqnarray}
    \int_{0}^3 \frac{ \rho (\zeta)  d \, \zeta}
    { \zeta + \zeta _0} = \int _0 ^3 \frac{\rho  d \zeta}{\zeta} +
%    \\
%    \nonumber
    \int _0 ^3 d\zeta \left[ \frac{\rho}{\zeta _0 + \zeta}
     - \frac{\rho}{\zeta} \right] 
     \\
     \nonumber
     = I - \zeta _0 \int _0 ^3 
	\frac{\rho d \zeta}{\zeta (\zeta + \zeta _0)}.
%    \\
%    \nonumber
\end{eqnarray}
To estimate
$\int _0 ^3 {d \zeta \rho(\zeta)}/[\zeta (\zeta + \zeta _0)]$,
we re-write it to show explicitly the most singular contribution
\begin{eqnarray}
    \int\!\! \frac{\rho d \zeta}{\zeta (\zeta \!+\! \zeta _0)} 
%\\
%\nonumber
%    =
%    \int_{0}^3 \frac{\left[ (\rho/\zeta) - \rho' (0)\right] d\zeta}
%    {\zeta + \zeta _0}
%	+ \rho' (0) \int_{0}^3 \frac{d\zeta }{\zeta + \zeta _0}
%\\
%\nonumber
    \!=\!
	\frac{2}{\sqrt{3}\pi} \ln \!\left(\! \frac{3}{\zeta _0}\! \right)
    \!+\!
    \int\!
	\frac{\left[ (\rho/\zeta) \!-\! \rho' (0)\right]\! d\zeta}
		{\zeta + \zeta _0}.
\end{eqnarray}
The remaining integral is finite in the limit
$\zeta_0 \rightarrow 0$,
as can be proven with the help of
Eq.~(\ref{eq::statedensity_lin}).
To evaluate it numerically we divide the integration interval into two
parts
\begin{eqnarray}
    \int_{0}^3 \frac{\left[ (\rho/\zeta) - \rho' (0)\right] d\zeta}
    {\zeta + \zeta _0} =
    \int^{\zeta^*}_0 \frac{\left[ (\rho/\zeta) - \rho' (0)\right] d\zeta}
    {\zeta + \zeta _0}
\\
\nonumber 
	+
    \int^{3}_{\zeta^*} \frac{\left[ (\rho/\zeta) - \rho' (0)\right] d\zeta}
    {\zeta + \zeta _0}.
\end{eqnarray}
The usefulness of this representation stems from the fact that 
numerically evaluated numerator has large error bars near zero $\zeta$.
Fortunately, for small
$\zeta^*$
and
$0 < \zeta < \zeta^*$
analytical approximation based on
Eq.~(\ref{eq::statedensity_lin}) 
can be used:
$\rho(\zeta) / \zeta  - \rho' (0) = {2 \zeta^{1/2}}/(3 \pi)$.
Integral over the interval between 
$\zeta^* = 0.2$
and 3, evaluated numerically, is found to be very small. Thus
\begin{eqnarray}
    \int_{0}^3 \frac{\left[ (\rho/\zeta) - \rho' (0)\right] d\zeta}
    {\zeta + \zeta _0} \approx
%    \\
%    \nonumber
%    \int_{0.2}^3 \frac{\left[ (\rho/\zeta) - \rho' (0)\right] d\zeta}
%    {\zeta + \zeta _0} +
%    \int_{0}^{0.2} \frac{\left[ (\rho/\zeta) - \rho' (0)\right] d\zeta}
%    {\zeta + \zeta _0} =
%    \\
%    \nonumber
%    =\int_{0.2}^3 \frac{\left[ (\rho/\zeta) - \rho' (0)\right] d\zeta}
%    {\zeta + \zeta _0} + \frac{4}{3 \pi}\int_{0}^{0.2} \frac{\zeta^{1/2} d\zeta}{(\zeta + \zeta _0)} \approx
%    \\
%    \nonumber
	\frac{4}{3 \pi} \sqrt{\zeta^*} \approx 0.2
\end{eqnarray}
for 
$\zeta^* = 0.2$.
Collecting all terms, we find
\begin{eqnarray}
%%%%%%%%%%%%%%%%%%%%%%%%%%%%%%%%%%%%%%%%%%%%%%%%%%
\label{eq::eta_approx_appendix}
%%%%%%%%%%%%%%%%%%%%%%%%%%%%%%%%%%%%%%%%%%%%%%%%%%
\eta(\zeta_0)
=
2I - \frac{2 \zeta _0}{\sqrt{3} \pi} \ln\left(\frac{3}{\zeta_0}\right)
+ O(\zeta_0),
\end{eqnarray}
where the retained terms are more singular than those which were ignored.
This expression is the basis for
Eq.~(\ref{eq::eta_approx}).
%Observe that we can re-write
%Eq.~(\ref{eq::eta_approx_appendix})
%as follows
%\begin{eqnarray}
%%%%%%%%%%%%%%%%%%%%%%%%%%%%%%%%%%%%%%%%%%%%%%%%%%%
%\label{eq::eta_approx_appendix2}
%%%%%%%%%%%%%%%%%%%%%%%%%%%%%%%%%%%%%%%%%%%%%%%%%%%
%\eta(\zeta_0)
%=
%2I - \frac{2 \zeta _0}{\sqrt{3} \pi} 
%	\left[ \ln\left(\frac{3}{\zeta_0}\right) + C \right]
%+ o(\zeta_0).
%\qquad
%\end{eqnarray}
%The order-of-unity constant $C$ is independent of $\zeta_0$ (as well as
%other parameters of the model) and, in principle, can be evaluated
%numerically. Its value modifies factor

\section{Mean field equations derivations}
%%%%%%%%%%%%%%%%%%%%%%%%%%%%%%%%%%%%%%%%%%%%%%%%%%
\label{app::MF_derivation}
%%%%%%%%%%%%%%%%%%%%%%%%%%%%%%%%%%%%%%%%%%%%%%%%%%

In this Appendix we will provide additional technical details for the mean
field equations derivations. Our goal is to differentiate
${\cal E} ^{\rm var}$,
defined by 
Eq.~(\ref{eq::E_var}),
over the variational parameters 
$\zeta_\perp$,
$\Tilde \zeta_0$,
and
$\Tilde \zeta_\perp$.
With this in mind, it is convenient to express
${\cal E} ^{\rm var}$
in the following manner
\begin{eqnarray}
%%%%%%%%%%%%%%%%%%%%%%%%%%%%%%%%%%%%%%%%%%%%%%%%%%
\label{eq::E_var_expand}
%%%%%%%%%%%%%%%%%%%%%%%%%%%%%%%%%%%%%%%%%%%%%%%%%%
{\cal E}^{\rm var}
=
({\cal N}_{\rm g} t)^{-1} \left(
	\langle \hat{H}^{\rm MF} \rangle
	+
	\langle \delta \hat{H} \rangle
	+
	\langle \hat{H}_{\rm HB} \rangle
\right)
+
\delta {\cal E}_{\rm m}.
\qquad
\end{eqnarray} 
In brief, this representation explicitly splits
$\langle \hat{H} \rangle$
into three terms: (i)~the mean field energy
$\langle \hat{H}^{\rm MF} \rangle$,
(ii)~the interaction term
$\langle \hat{H}_{\rm HB} \rangle$,
and (iii)~all other contributions
$\langle \delta \hat{H} \rangle$.
The Hamiltonian in (iii) is bilinear in single-electron operators 
\begin{eqnarray}
\delta \hat{H} = \sum _{\mathbf{q}}
    \hat\Phi_{\bf q}^\dag
    \delta \hat{\cal H}
    \hat\Phi_{\bf q}^{\vphantom{\dag}},
\end{eqnarray} 
its associated matrix
$\delta \hat{\cal H}$
equals to
\begin{eqnarray} 
\delta \hat{\cal H}
=
t \left( 
\begin{matrix}
	\delta \zeta_0 & 0 & -i \delta \zeta_\perp  & 0 \\
0  & \delta \zeta_0 & 0 & i \delta \zeta_\perp \\
i \delta \zeta_\perp & 0 & -\delta \zeta_0 & 0\\
0 & -i \delta \zeta_\perp & 0 & -\delta \zeta_0 
    \end{matrix} \right),
\end{eqnarray}
where
$\delta \zeta_0 = \zeta_0 - \tilde \zeta_0$,
and
$\delta \zeta_\perp = \zeta_\perp - \tilde \zeta_\perp$.

Let us differentiate
${\cal E}^{\rm var}$
over 
$\tilde \zeta_\perp$.
The term
$\delta {\cal E}_{\rm m}$
is independent of
$\tilde \zeta_\perp$,
thus the corresponding derivative vanishes. The Hellmann-Feynman theorem
allows us to establish that 
\begin{eqnarray}
%%%%%%%%%%%%%%%%%%%%%%%%%%%%%%%%%%%%%%%%%%%%%%%%%% 
\label{eq::HlFn}
%%%%%%%%%%%%%%%%%%%%%%%%%%%%%%%%%%%%%%%%%%%%%%%%%% 
\frac{\partial\langle\hat{H}^{\rm MF}\rangle}{\partial\Tilde\zeta_\perp}
=
2 t \langle S^{y} \rangle {\cal N}_{\rm g}.
\end{eqnarray} 
Here we used the relation 
$\langle S ^{y}_{{\bf R}_A} \rangle = 
- \langle S ^{y}_{{\bf R}_B} \rangle = \langle S^{y} \rangle$
which connects magnetization projections on the two sublattices.

To calculate the derivatives for two other terms in
Eq.~(\ref{eq::E_var_expand})
we can write explicit expressions for them
\begin{eqnarray}
\langle \delta \hat{H} \rangle
=
2 t \left(
	\delta \zeta_0 \langle S_z \rangle
	+
	\delta \zeta_\perp \langle S_y \rangle
\right) {\cal N}_{\rm g},
\\
\langle \hat{H}_{\rm HB} \rangle
=
\frac{U}{2}\left(1 - \langle S_z\rangle^2 - \langle S_y\rangle^2\right){\cal N}_{\rm g}.
\end{eqnarray} 
Therefore
\begin{eqnarray}
\frac{\partial \langle \delta \hat{H} \rangle}{\partial \tilde \zeta_\perp}
=
2 t \left(
	 - \langle S_y \rangle
	 +
	 \delta \zeta_\perp
	 \frac{\partial \langle S_y \rangle} {\partial \tilde \zeta_\perp}
\right) {\cal N}_{\rm g},
\\
\frac{\partial \langle \hat{H}_{\rm HB} \rangle }
	{\partial \tilde \zeta_\perp}
=
- U \langle S_y \rangle
	 \frac{\partial \langle S_y \rangle} {\partial \tilde \zeta_\perp}
{\cal N}_{\rm g}.
\end{eqnarray} 
Collecting all contributions, we obtain
\begin{eqnarray}
\frac{\partial {\cal E}^{\rm var}}{\partial \tilde \zeta_\perp}
=
 \left(
	2 \delta \zeta_\perp - \frac{U}{t} \langle S_y \rangle
\right) \frac{\partial \langle S_y \rangle} {\partial \tilde \zeta_\perp}.
\end{eqnarray} 
Equating the latter with zero one recovers
Eq.~(\ref{eq::perpdiff}).
Two other mean field equations are derived using similar tactics.

Finally, we want to outline the derivation of
expression~(\ref{eq::cantaxisy})
for
$\langle S^y \rangle$.
The most convenient approach is to use 
Eq.~(\ref{eq::HlFn}).
The mean field energy is
\begin{eqnarray}
%%%%%%%%%%%%%%%%%%%%%%%%%%%%%%%%%%%%%%%%%%%%%%%%%%
\label{eq::E_MF}
%%%%%%%%%%%%%%%%%%%%%%%%%%%%%%%%%%%%%%%%%%%%%%%%%%
\langle \hat{H}^{\rm MF} \rangle
=
- {\cal N}_{\rm g} t
\int _0 ^3 d \zeta \, \rho
	\sqrt{\Tilde{\zeta} _{\perp} ^2 +
    	\left( \Tilde{\zeta} _0 - \zeta \right) ^2}
-
\\
\nonumber 
{\cal N}_{\rm g} t
\int _0 ^3 d \zeta \, \rho
    \sqrt{\Tilde{\zeta}_{\perp}^2+\left(\Tilde{\zeta}_0+\zeta\right)^2}.
\quad
\end{eqnarray}
Substituting this expression into
Eq.~(\ref{eq::HlFn}),
one derives
\begin{eqnarray}
%%%%%%%%%%%%%%%%%%%%%%%%%%%%%%%%%%%%%%%%%%%%%%%%%%
\label{eq::Sy}
%%%%%%%%%%%%%%%%%%%%%%%%%%%%%%%%%%%%%%%%%%%%%%%%%%
2\langle S^y \rangle = 
- \Tilde \zeta_\perp
\int _0 ^3 
	\frac{\rho (\zeta) d \zeta}
	{\sqrt{\Tilde\zeta_\perp^2 + \left(\Tilde\zeta_0 - \zeta\right)^2}}
-
\\
\nonumber 
\Tilde \zeta_\perp
\int _0 ^3 \frac{\rho(\zeta) d \zeta}
		{\sqrt{\Tilde{\zeta} _{\perp} ^2 +
    \left( \Tilde{\zeta} _0 + \zeta \right) ^2}}.
\end{eqnarray} 
The derivation steps discussed in
Appendix~\ref{app::gap_derivation}
allows one to recover
Eq.~(\ref{eq::cantaxisy})
from
Eq.~(\ref{eq::Sy}).

%\bibliographystyle{apsrevlong_no_issn_url}
%\bibliography{Graph_magn}

\end{document}